\documentclass[aps,prresearch,reprint,amsmath,amssymb,superscriptaddress,floatfix]{revtex4-2}

\usepackage{graphicx}
\usepackage{bm}
\usepackage{siunitx}
\usepackage{microtype}
\usepackage{hyperref}
\hypersetup{colorlinks=true,citecolor=blue,urlcolor=blue,linkcolor=blue}
\begin{document}

\title{Large Language Model Steering of Acousto-Optic Tweezer Arrays}

\author{Cheng Sheng}
\email{shengcheng@ccnu.edu.cn}
\affiliation{Department of Physics, Huazhong Normal University, Wuhan 430079, China}
\author{Kai-xuan Xiao}
\affiliation{Department of Physics, Huazhong Normal University, Wuhan 430079, China}
\author{Yin Zhang}
\affiliation{Department of Physics, Huazhong Normal University, Wuhan 430079, China}
\author{Wen-hao Wang}
\affiliation{Department of Physics, Huazhong Normal University, Wuhan 430079, China}
\author{Feng-yi Zhang}
\affiliation{Department of Physics, Huazhong Normal University, Wuhan 430079, China}
\author{Xu-xing Geng}
\affiliation{Innovation Academy for Precision Measurement Science and Technology, Chinese Academy of Sciences, Wuhan 430071, China}
\author{Guang-ming Huang}
\affiliation{Department of Physics, Huazhong Normal University, Wuhan 430079, China}
\author{Gao-xiang Li}
\email{gaox@ccnu.edu.cn}
\affiliation{Department of Physics, Huazhong Normal University, Wuhan 430079, China}

\begin{abstract}
Controlling quantum hardware requires choosing and revising experimental strategies as nonlinear response, coupling, and drift limit calibration. We experimentally investigate a general-purpose large language model (LLM) as a human-in-the-loop strategy layer for a neutral-atom tweezer platform. Through step-by-step interaction with experimental feedback, the LLM guides a transition from multiplicative row--column correction to a drift-corrected local response-matrix method when separable feedback stalls. Deterministic software performs image analysis and numerical updates, while an experimentalist verifies and authorizes each waveform. For a $10\times10$ crossed-acousto-optic-deflector array, this workflow reduces the Gaussian-peak coefficient of variation from $20.30\%$ to $1.57\%$ in 30 candidate-update rounds. The experiment demonstrates feedback-informed strategy selection on a physical quantum-control subsystem, rather than direct generation of control parameters alone. These results highlight the potential of LLMs to guide control optimization across diverse quantum platforms by adaptively formulating and refining experimental strategies.
\end{abstract}

\maketitle

\begin{figure*}[t!]
\centering
\includegraphics[width=0.98\textwidth]{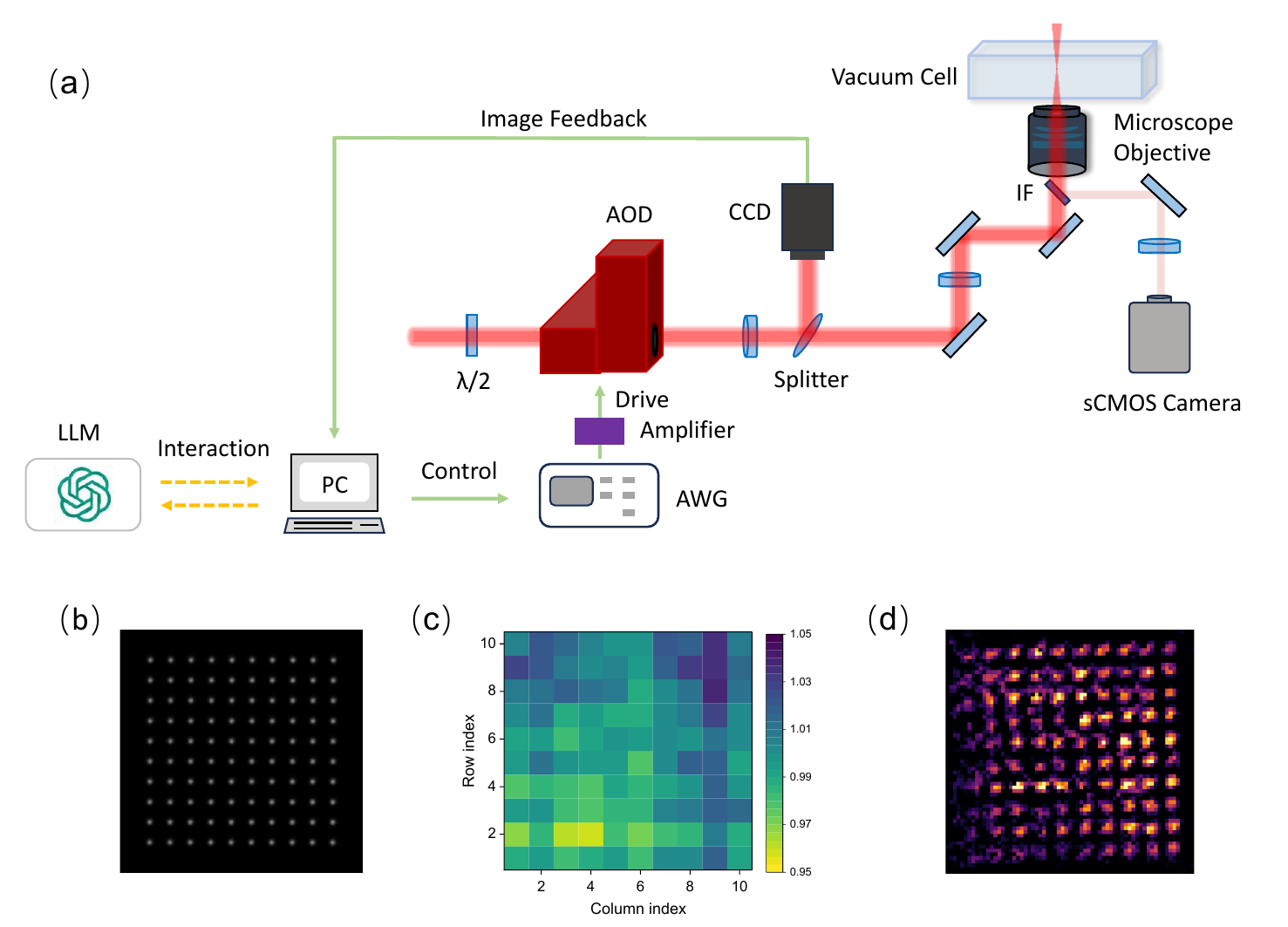}
\caption{LLM-guided optical control on a neutral-atom platform. (a) An $852$-nm laser generates tweezers for $^{87}$Rb atoms. A crossed AOD pair is driven by an arbitrary waveform generator (AWG) through RF amplifiers. A 1:99 beamsplitter directs $1\%$ of the optical power to the feedback CCD and $99\%$ through an interference filter (IF) and microscope objective toward the vacuum cell; a sCMOS camera records the science image. The LLM exchanges measured states and proposed strategies with the computer through a human-verified interface. (b) Measured $10\times10$ tweezer intensity array on the feedback camera. (c) Normalized fitted-peak map at the optimized setting. (d) Representative fluorescence image of the trapped-atom array recorded by the science camera.}
\label{fig:platform}
\end{figure*}

Programmable optical-tweezer arrays are a leading platform for neutral-atom quantum computation and simulation~\cite{Saffman2010,Browaeys2020,Kaufman2021}. Recent milestones encompass midcircuit measurement, dual-species Rydberg operations, quantum-information dynamics, AI-assisted array assembly, thousands of coherent atomic qubits, and building blocks for fault-tolerant computation~\cite{Graham2023,Anand2024,Liang2025,Lin2025,Chiu2025,Manetsch2025,Bluvstein2026,Xiang2026,Senoo2026}. These advances place increasing demands on the classical hardware that sets trap depths and positions, transports atoms, and addresses selected sites. Spatial nonuniformity produces site-dependent potentials and light shifts, increasing the calibration burden for state preparation and coherent control.

Acousto-optic deflectors (AODs) are particularly useful because radio-frequency (RF) amplitudes and frequencies dynamically control tweezer depths and positions without mechanical motion. This programmability supports atom-by-atom assembly, coherent gates, enhanced loading, efficient rearrangement, and mixed-species arrays~\cite{Endres2016,Sheng2018,Brown2019,Sheng2021,Sheng2022}. Recent developments extend AOD control to model-based synthesis of $50\times50$ arrays~\cite{Mittenbuehler2025} and astigmatism-free three-dimensional trajectories~\cite{Lu2025}. Calibration nevertheless remains difficult because diffraction efficiency varies with frequency, multifrequency amplification produces intermodulation, and coupled optical and RF responses drift. Sun \textit{et al.} combined multistage random searches over candidate phase configurations with intensity-error-based amplitude feedback, achieving an intensity error of $\sigma=0.02$ in a $10\times10$ AOD beam array after approximately four minutes of closed-loop optimization~\cite{Sun2024}. Their amplitude correction uses a calibrated per-frequency intensity response, while a progressively narrowed random search optimizes the phases. The present work extends this approach at the strategy level: experimental feedback informs not only parameter updates but also the choice of control model. In particular, the LLM can replace a separable description with a measured cross-coupled response matrix and recommend re-identification when drift invalidates that matrix. This adaptability addresses a limitation of retaining a prescribed feedback model as the experimental operating regime changes.

Artificial intelligence already contributes to quantum-device design, calibration, control, and error correction~\cite{Alexeev2025}. Cold-atom demonstrations include Gaussian-process optimization of Bose--Einstein-condensate production~\cite{Wigley2016}, neural-network optimization of a 63-parameter magneto-optical-trap sequence~\cite{Barker2018}, and reinforcement-learning policies that adapt to experimental perturbations~\cite{Reinschmidt2024}. These approaches typically optimize within a specified representation and objective. LLMs offer a complementary interface: they can propose candidates from a natural-language optimization history~\cite{Yang2024} and connect laboratory context to executable procedures. Recent work includes agent-based superconducting experiments~\cite{Cao2025}, LLM-assisted qubit-control routines~\cite{Li2026}, safety-gated trapped-ion control code~\cite{Wang2026}, and the generation of quantum-experiment programs and control ansatzes~\cite{Arlt2026,Zhao2026}. Here we investigate a distinct role: using residual patterns, failed updates, and drift to decide when the mathematical \emph{representation} of an experimental control problem should change.

We test this role on the optical control subsystem of a neutral-atom apparatus using OpenAI GPT-5.5 Thinking through ChatGPT Plus. In step-by-step interactions, an experimentalist supplied images, fitted intensity matrices, the accepted waveform, and the preceding update's outcome [Fig.~\ref{fig:platform}(a)]. The LLM recommended a model, update scale, and next action; deterministic software performed fitting, matrix algebra, and waveform generation, and the experimentalist authorized every hardware update. The resulting strategy switched from separable row--column feedback to drift-corrected local system identification, reducing the intensity CV from $20.30\%$ at R0 to a minimum of $1.57\%$ in the sequence R0--R30. A round denotes evaluation of a proposed operating waveform; response-matrix perturbations and baseline checkpoints are additional calibration measurements, detailed in the Supplemental Material. We therefore report the observed strategy and convergence trajectory, not a demonstrated reduction in measurement cost relative to conventional feedback. This experiment establishes a practical link between language-based strategy selection and quantitative hardware control, illustrating how an LLM can help revise a calibration model in response to evidence from a working quantum platform.

\begin{figure}[t!]
\centering
\includegraphics[width=0.98\columnwidth]{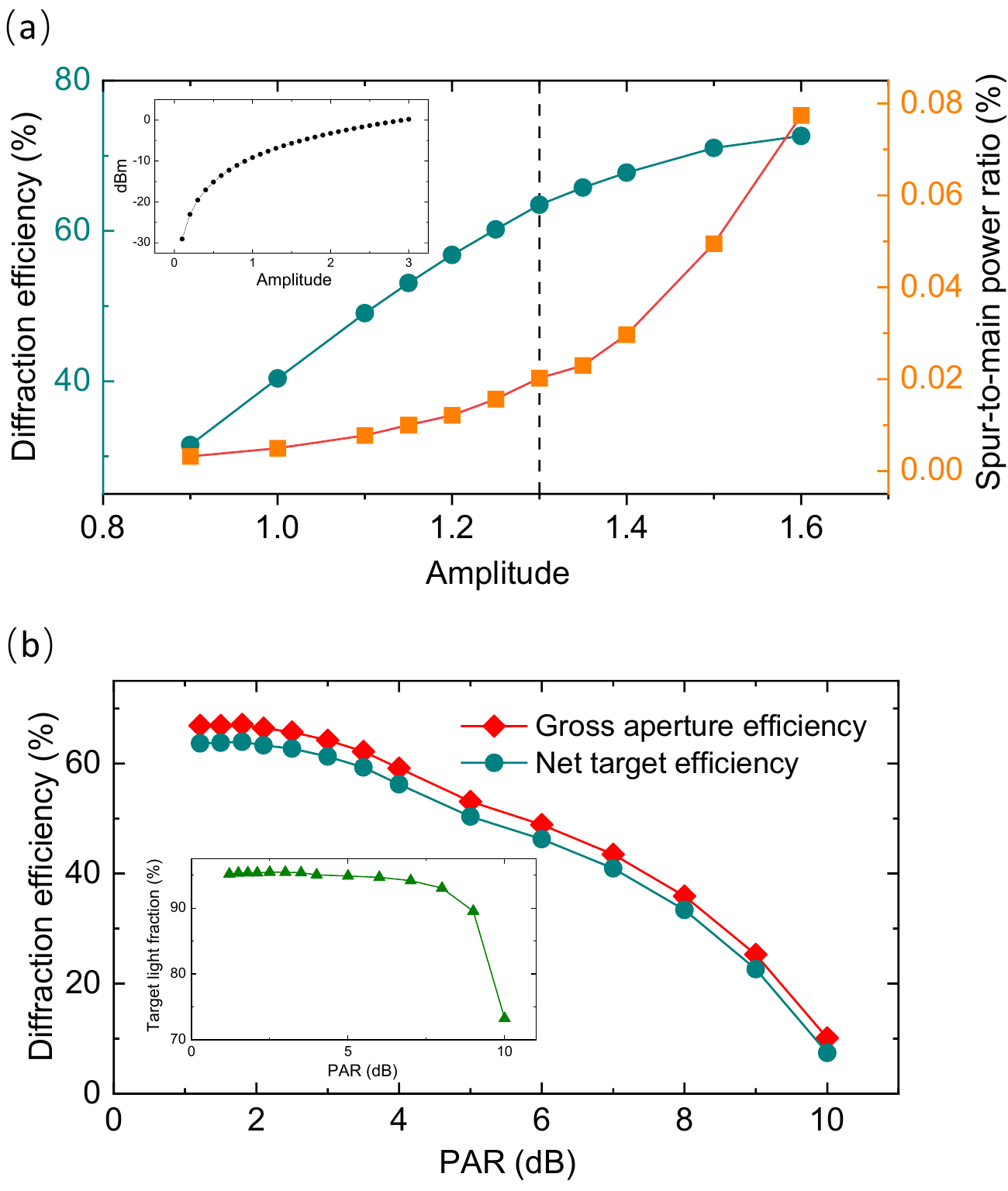}
\caption{Choice of the RF operating regime. (a) Diffraction efficiency (teal circles, left axis) and spur-to-main RF-power ratio (orange squares, right axis) versus normalized multifrequency amplitude. The dashed line marks $A=1.30$, where diffraction is high but the spur ratio has not yet entered its rapid-growth regime; the inset calibrates normalized AWG amplitude to RF power in dBm. (b) Gross aperture efficiency (red diamonds) and net target efficiency (teal circles) for 14 phase conditions ordered by peak-to-average power ratio (PAR). Both efficiencies deteriorate with increasing PAR; the inset shows the accompanying loss of target-array power fraction. The scan defines a low-PAR, low-spur operating region before the LLM-guided amplitude optimization.}
\label{fig:rf}
\end{figure}

To implement this strategy-guided workflow, we first characterize the RF-to-optical response and select a stable operating regime for the crossed-AOD array. Each AOD axis is driven by ten tones from $95$ to $104~\mathrm{MHz}$. We denote the two channels by CH1 and CH2; CH1 predominantly controls columns and CH2 rows. The waveform is
\begin{equation}
V_{\alpha}(t)=\sum_{m=1}^{10}A_{\alpha m}\cos(2\pi f_m t+\phi_{\alpha m}),
\qquad \alpha\in\{1,2\}.
\label{eq:waveform}
\end{equation}
The control vector therefore contains 20 amplitudes, while the observed vector contains 100 optical intensities with weak nonlocal dependence on every control component. We first selected a common initial amplitude and a phase set for each channel, before allowing individual amplitudes to change in the uniformity loop. This separated the choice of an RF operating regime from the correction of site-dependent intensity errors.

For amplitude selection, all tones were assigned the same amplitude $A$, and the composite signal was scanned at fixed reference phases. Figure~\ref{fig:rf}(a) compares optical diffraction efficiency with the spur-to-main RF-power ratio, while its inset calibrates the normalized AWG amplitude against measured RF power. Increasing $A$ initially improves the useful optical power, but beyond approximately $A=1.30$ the spur ratio grows rapidly. Maximizing diffraction efficiency alone would therefore move the system toward stronger nonlinear distortion and less reliable amplitude control. We selected $A=1.30$ as a compromise between useful diffraction and a low-spur operating regime, rather than the slightly higher efficiency available at $A=1.35$ or above.

Phase selection was then examined at the selected common amplitude. Relative phases change the peak excursions of the composite waveform even when the tone amplitudes are unchanged, and thus affect its susceptibility to amplifier nonlinearity. Figure~\ref{fig:rf}(b) compares 14 phase conditions through their waveform peak-to-average power ratio (PAR), with phase sets specified independently for CH1 and CH2. Both gross aperture efficiency and net target efficiency decrease as PAR increases, accompanied by a reduced fraction of aperture power in the intended array. Low PAR therefore served as an initial screening criterion, followed by CCD evaluation of the actual array: similar PAR and efficiency do not guarantee the same spatial intensity pattern. The selected low-PAR phase configuration was retained for the uniformity sequence, during which the phases were held fixed and only the 20 amplitudes were updated. This initialization supplies the subsequent feedback with a reproducible, controllable operating point; the PAR definition and supplementary phase comparison are given in the Supplemental Material.

\begin{figure*}[t!]
\centering
\includegraphics[width=0.99\textwidth]{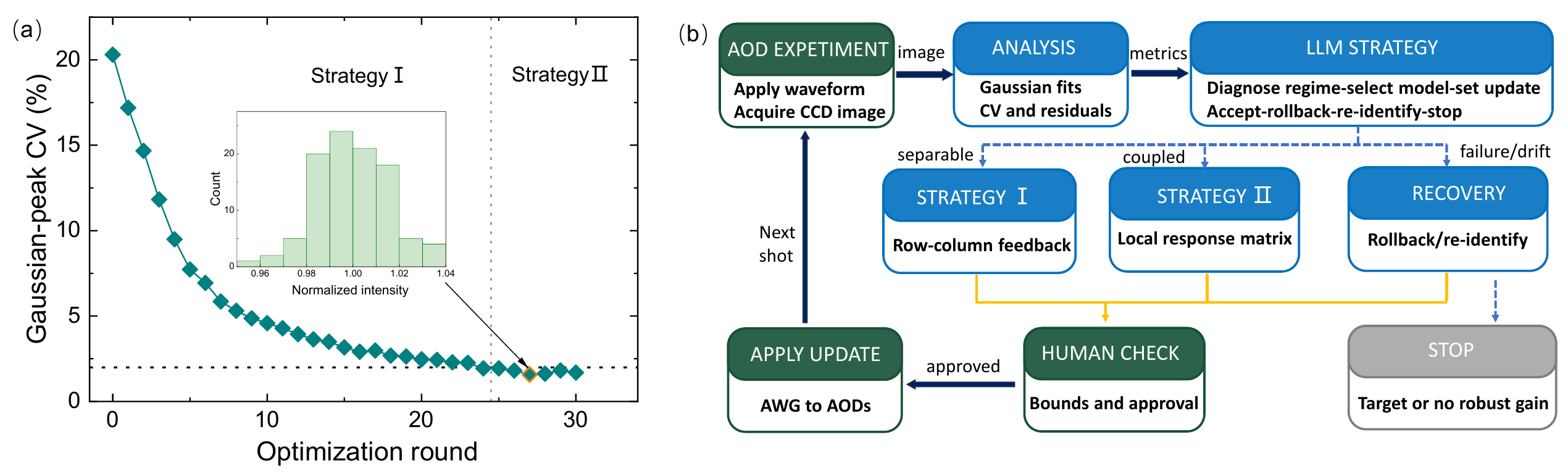}
\caption{Experimental trajectory and step-by-step LLM control workflow. (a) The Gaussian-peak CV decreases under Strategy I, separable row--column feedback, before promotion to Strategy II, a measured $100\times20$ local response matrix. The orange diamond identifies the response-matrix state whose 100-site normalized-intensity distribution is shown in the inset; the lowest plotted CV occurs at R27, with a reported minimum of $1.57\%$, whereas the final plotted state R30 has a CV of approximately $1.7\%$. (b) At each round the experiment supplies an image, deterministic analysis returns CV and residual metrics, and GPT-5.5 Thinking selects a strategy and proposes the next update. A human verifies bounds and authorizes the AWG waveform; failure or drift triggers rollback or re-identification, while a reached target or absence of robust gain terminates the loop.}
\label{fig:trajectory}
\end{figure*}

The measurement pipeline was fixed before comparing control states. After same-day pixelwise background subtraction, each site was fitted in a fixed $17\times17$-pixel region, and the fitted optical peak was used to calculate the 100-site coefficient of variation (CV). The Gaussian model, CV definition, multi-frame averaging procedure, and PAR definition are given in the Supplemental Material. This deterministic pipeline is independent of the LLM and prevents a change in analysis from masquerading as an improvement in hardware control. The measured optimization trajectory and the step-by-step interaction logic are summarized together in Fig.~\ref{fig:trajectory}.

The initial intensity error had a strong row-column structure. Given the normalized peak matrix $M_{ij}=I_{ij}/\bar I$, the LLM-guided first strategy compressed 100 observations into the row and column means
\begin{equation}
R_i=\frac{1}{10}\sum_j M_{ij}, \qquad C_j=\frac{1}{10}\sum_i M_{ij},
\end{equation}
and used the multiplicative feedback law
\begin{align}
A^{\rm new}_{{\rm CH2},i}&=A^{\rm old}_{{\rm CH2},i}R_i^{-\gamma_2/2},\\
A^{\rm new}_{{\rm CH1},j}&=A^{\rm old}_{{\rm CH1},j}C_j^{-\gamma_1/2}.
\label{eq:rowcolumn}
\end{align}
The factor $1/2$ follows from the leading quadratic dependence of optical intensity on RF amplitude. The relaxation factors $\gamma_{1,2}$ convert the formula from a full correction into a bounded experimental step. They reached approximately $0.4$ while the error was large, then decreased and became channel dependent as the array approached uniformity. This low-dimensional model reduced the CV from $20.30\%$ at R0 to approximately $3\%$ around R16. A small rebound at R17 prompted rollback and reduced gains. The subsequent row--column updates brought the CV toward $2\%$ near R24--R25, where progress slowed.

The stalled improvement was accompanied by residuals that were no longer separable into row and column envelopes. The LLM therefore changed the control representation from two marginal vectors to a local multiple-input multiple-output model. With $u_k=\ln A_k$ and $y_p=\ln M_p$, the measured response is
\begin{equation}
\Delta \bm y=J\Delta\bm u, \qquad
J_{pk}=\frac{\partial\ln M_p}{\partial\ln A_k}.
\label{eq:jacobian}
\end{equation}
The $100\times20$ Jacobian was obtained by symmetric perturbations of each tone. The update minimized
\begin{equation}
\|J\Delta\bm u+\bm y\|_2^2+\lambda^2\|\Delta\bm u\|_2^2
\label{eq:ridge}
\end{equation}
and was clipped by a per-tone relative-amplitude cap. This construction turned the LLM proposal into an auditable control law rather than a list of language-generated amplitudes.

The response-matrix stage in Fig.~\ref{fig:trajectory}(a) lowers the CV below the preceding row--column plateau. The lowest plotted point is R27, with a reported CV of $1.57\%$; R28--R30 remain close to this level, with the final R30 value approximately $1.7\%$. During local-model refinement, the LLM-guided response to drift was to re-identify the model rather than continue using an outdated correction direction. Each tone was perturbed by $\pm0.4\%$, and repeated baseline checkpoints were used to remove time-dependent drift from the finite differences. Regularization and per-tone amplitude caps constrained the numerical updates (see the Supplemental Material).

The small variations near the end of the displayed trajectory do not establish a fundamental uniformity limit. Plausible limitations include image-fit repeatability, slow optical and RF drift, and error in a sequentially measured local Jacobian. Small corrections may be obscured by these variations, whereas larger extrapolations can leave the regime in which the local model is valid. These mechanisms were not independently quantified at the best plotted state. The minimum of $1.57\%$ should therefore be interpreted as the best observed CV in Fig.~\ref{fig:trajectory}(a), not a statistically established noise floor or a fundamental AOD limit.

The distinction between procedure automation and strategy formation is central. A plausible explanation for the LLM's usefulness is its ability to connect established ideas from optical diagnostics, feedback control, regularization, and drift compensation with the evolving experimental record. The present experiment does not isolate the contribution of pretraining or establish superiority over an expert-designed adaptive controller. It does show a concrete sequence: row--column structure motivated a low-dimensional controller; a plateau with nonseparable residuals motivated a measured local Jacobian; a failed extrapolation triggered rollback; and day-to-day drift prompted re-identification with stronger regularization and smaller steps. Compared with the per-frequency amplitude correction and random phase search of Ref.~\cite{Sun2024}, the strategy-level advantage is the ability to revise the control representation when residual cross-coupling or drift becomes important. A measured $100\times20$ Jacobian includes the response of every observed site to every controlled tone, and regularized, bounded updates use that information without assuming independent intensity responses. The improvement beyond the row--column plateau demonstrates the value of this model change within our apparatus; it does not by itself constitute a matched cross-apparatus benchmark. The contribution is the experimental realization of this feedback-informed composition of established methods, rather than the invention of a new numerical optimizer.

The optimized observable is optical intensity, not a measured qubit fidelity or many-body observable, so Fig.~\ref{fig:trajectory} does not establish a numerical gate-fidelity improvement. It does, however, close the LLM loop around hardware that defines site-dependent potentials in a neutral-atom processor. The same layered architecture can be extended by replacing camera intensity with atomic observables and by exposing only safety-bounded tools: loading probability for array assembly, survival and heating during transport, differential light shifts and Rabi-frequency maps for single-qubit calibration, blockade leakage and parity or Bell-state fidelity for entangling gates, and composite objectives that trade fidelity against atom loss and experimental time. At larger scale, an LLM could coordinate these coupled calibration layers, choose when to reuse or re-identify local models, and hand numerically precise subproblems to deterministic optimizers while maintaining provenance and human authorization.

In summary, OpenAI GPT-5.5 Thinking steered a real, drifting quantum-control subsystem through the step-by-step sequence R0--R30 by selecting, testing, and revising its mathematical model. The resulting hierarchy of separable feedback, local system identification, rollback, and drift-aware re-identification reduced the intensity CV of a $10\times10$ crossed-AOD array to a minimum of $1.57\%$ within the sequence ending at R30. The broader opportunity is not to replace control theory with generated text, but to use an LLM as a strategy layer that decides which verified controller and which physical abstraction should govern the next experiment. By connecting experimental feedback to adaptive strategy selection, this work provides a proof of principle for LLM-assisted calibration of neutral-atom quantum processors and motivates its extension to other quantum-control tasks.

This work was supported by the Natural Science Foundation of Hubei Province (Grant No. 2025AFB851), by the National Natural Science Foundation of China (Grant No. 12374330), and by the Fundamental Research Funds for the Central Universities (Grant Nos. 30106260175 and JC2026TS-003).

C. S. and K.-X. X. contributed equally to this work.

\clearpage
\onecolumngrid
\begin{center}
{\large\bfseries Supplemental Material for Large Language Model Steering of Acousto-Optic Tweezer Arrays}\par
\vspace{0.5em}

{Cheng Sheng$^{1}$, Kai-xuan Xiao$^{1}$, Yin Zhang$^{1}$, Wen-hao Wang$^{1}$, Feng-yi Zhang$^{1}$, Xu-xing Geng$^{2}$, Guang-ming Huang$^{1}$, and Gao-xiang Li$^{1}$}\par
\vspace{0.3em}
{\itshape $^{1}$Department of Physics, Huazhong Normal University, Wuhan 430079, China}\par
{\itshape $^{2}$Innovation Academy for Precision Measurement Science and Technology, Chinese Academy of Sciences, Wuhan 430071, China}
\par\vspace{0.2em}
{\small shengcheng@ccnu.edu.cn; gaox@ccnu.edu.cn}
\end{center}
\vspace{0.8em}
\twocolumngrid
\small
\raggedbottom

\setcounter{equation}{0}
\renewcommand{\theequation}{S\arabic{equation}}
\renewcommand{\theHequation}{S\arabic{equation}}
\setcounter{figure}{0}
\renewcommand{\thefigure}{S\arabic{figure}}
\renewcommand{\theHfigure}{S\arabic{figure}}
\setcounter{section}{0}
\renewcommand{\thesection}{S\arabic{section}}
\renewcommand{\theHsection}{S\arabic{section}}

\section{Control hardware and waveform definitions}
\label{sec:hardware}

An $852$-nm laser generates the tweezer potential for $^{87}$Rb atoms. Two orthogonal AODs (DTXSXY-400-850) form the $10\times10$ array and are driven by an M4x.6622-x4 AWG through ZHL-1-2W-S+ RF amplifiers. A 1:99 beamsplitter sends $1\%$ of the power to the feedback CCD and $99\%$ through an interference filter (IF) toward the vacuum cell; a 400BSI V3 sCMOS camera records the science images. Each RF channel contains ten frequencies stored in ascending order from $95$ to $104~\mathrm{MHz}$. CH1 predominantly controls columns: $104~\mathrm{MHz}$ maps to the leftmost CCD column and $95~\mathrm{MHz}$ to the rightmost. CH2 predominantly controls rows: $104~\mathrm{MHz}$ maps to the top CCD row and $95~\mathrm{MHz}$ to the bottom. The waveform on channel $\alpha$ is
\begin{equation}
V_{\alpha}(t)=\sum_{m=1}^{10} A_{\alpha m}\cos(2\pi f_m t+\phi_{\alpha m}).
\end{equation}
The initial common amplitude was $A=1.30$ for each tone. The uniformity sequence held the selected low-PAR phase configuration fixed and varied only the 20 amplitudes, as described in Secs.~\ref{sec:efficiency} and \ref{sec:phaseselection}.

For a multifrequency channel with instantaneous power $p_\alpha(t)\propto V_\alpha^2(t)$, the peak-to-average power ratio (PAR) is
\begin{equation}
\begin{aligned}
\mathrm{PAR}_\alpha&=10\log_{10}\!\left[\frac{\max_t p_\alpha(t)}{\langle p_\alpha(t)\rangle_t}\right]\\
&=20\log_{10}\!\left[\frac{\max_t|V_\alpha(t)|}{V_{\alpha,\mathrm{rms}}}\right].
\end{aligned}
\end{equation}

\section{Image processing and uniformity metric}
\label{sec:imageprocessing}

A same-day background image was subtracted pixel by pixel. A fixed table of 100 spot centers then defined $17\times17$-pixel regions of interest. Each region was fitted to an axis-aligned two-dimensional Gaussian with a constant background,
\begin{equation}
G(x,y)=B+I\exp\!\left[-\frac{(x-x_0)^2}{2\sigma_x^2}-\frac{(y-y_0)^2}{2\sigma_y^2}\right].
\end{equation}
The Gaussian peak $I$ was used for uniformity. The Gaussian volume $2\pi I\sigma_x\sigma_y$ was used only when estimating optical power. For $N=100$ sites, the population coefficient of variation is
\begin{equation}
\bar I=\frac{1}{N}\sum_{p=1}^{N}I_p,\qquad
\mathrm{CV}=\frac{\sqrt{N^{-1}\sum_{p=1}^{N}(I_p-\bar I)^2}}{\bar I}.
\end{equation}
For multiple images of one state, corresponding fitted peak values were averaged site by site before calculating the 100-site CV. This is distinct from averaging the CV values of individual images. The equation above defines CV as a dimensionless ratio; its numerical value is multiplied by 100 when expressed as a percentage. Row and column summaries were obtained from the ten row-averaged and ten column-averaged intensities; they served as diagnostic inputs to the feedback rather than replacing the 100-site CV.

\section{Diffraction efficiency and amplitude selection}
\label{sec:efficiency}

Let $P_{\rm in}$ be the input optical power, $P_{\rm off}$ the aperture power with the RF off, $P_{\rm on}$ the aperture power with the RF on, and $P_{\rm bg}$ the power-meter background. The target fraction $r_{\rm aperture}$ is the sum of the 100 fitted Gaussian volumes divided by the total background-subtracted CCD signal inside the aperture. The net target power and net diffraction efficiency are
\begin{align}
P_{\rm target}&=r_{\rm aperture}(P_{\rm on}-P_{\rm off}),\\
\eta_{\rm net}&=\frac{P_{\rm target}}{P_{\rm in}-P_{\rm bg}}.
\end{align}
The final mean RF amplitude was $A=1.30$. Although an earlier scan suggested that $A=1.35$ could give slightly higher efficiency, the larger amplitude also produced stronger intermodulation and less stable uniformity feedback. The experiment therefore treated efficiency as a constraint and selected the more controllable operating point.

\section{Phase selection and fixed-PAR comparison}
\label{sec:phaseselection}

\begin{figure}[!htb]
\centering
\includegraphics[width=0.98\columnwidth]{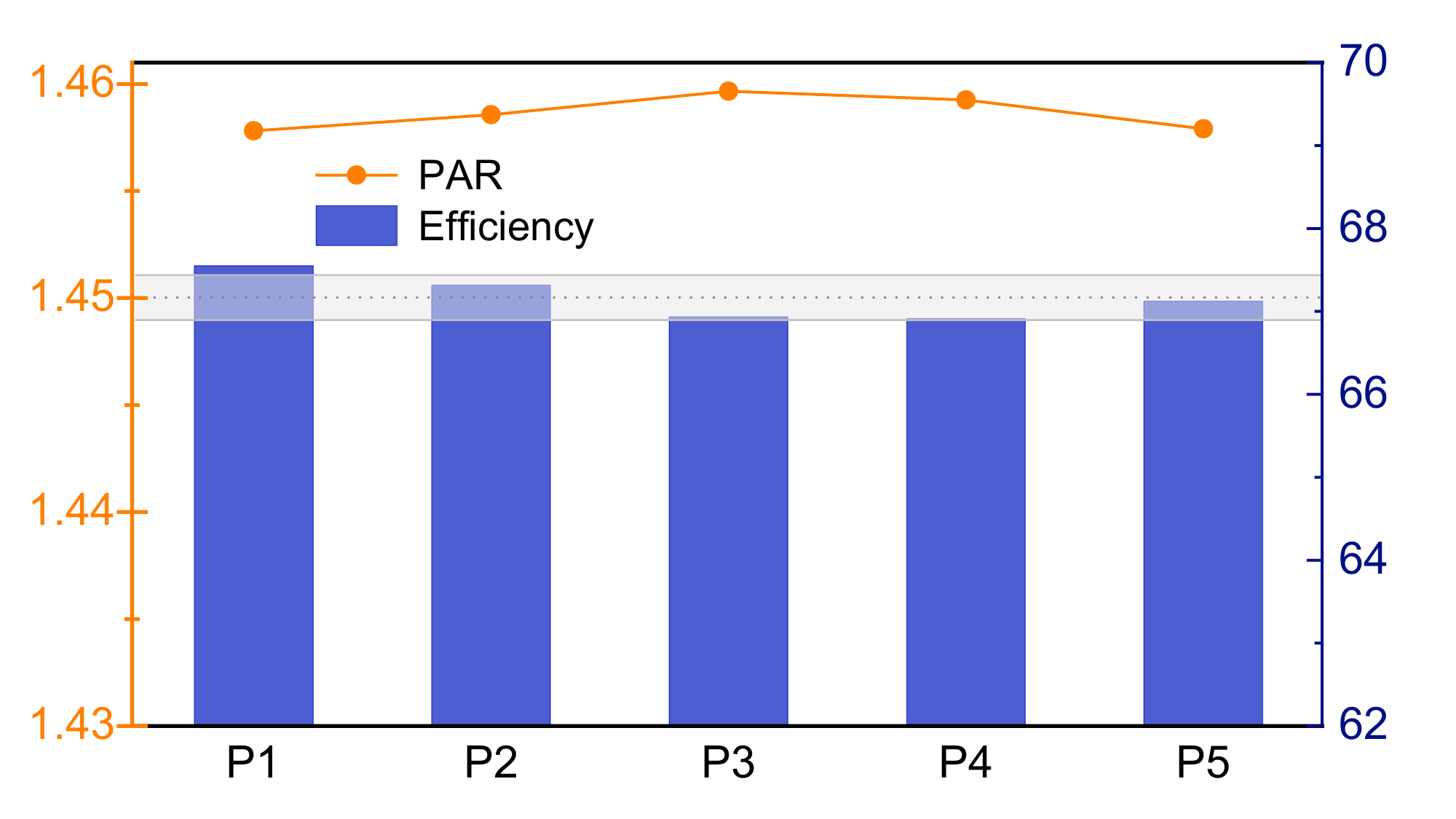}
\caption{Phase control at nearly fixed PAR. The orange markers show the PAR of five phase configurations, labeled P1--P5 for this comparison, and the blue bars show their measured diffraction efficiencies. The narrow efficiency spread supports the use of PAR to screen phase settings while holding the mean drive amplitude fixed. Spatial uniformity was still verified independently on the CCD.}
\label{fig:phasecontrol}
\end{figure}

At the common amplitude $A=1.30$, the 14-condition phase scan in Fig.~\ref{fig:rf}(b) identified a low-PAR operating region. The selected phase configuration was chosen using both the RF waveform properties and the measured CCD intensity distribution. It was labeled P2 in the experimental record; P2 is a configuration identifier, not a numerical PAR value or a distinct optimization algorithm. The additional fixed-PAR comparison separated waveform peak statistics from the detailed arrangement of phases. Figure~\ref{fig:phasecontrol} shows five phase settings with nearly identical PAR. Their diffraction efficiencies agree within the observed spread, supporting PAR as a rapid screening variable for the amplifier operating regime. This does not imply that PAR uniquely predicts the spatial uniformity of the $10\times10$ array: phase sets with similar PAR can produce different site-resolved patterns after propagation through the crossed-AOD system.

\section{Step-by-step LLM interaction}
\label{sec:llminteraction}

The campaign used OpenAI GPT-5.5 Thinking through ChatGPT Plus in a step-by-step interaction mode. At each decision point the human operator provided the current uniformity metrics, row and column summaries, the accepted RF state, and relevant diagnostic plots or failed updates. The LLM returned a proposed model, its feedback scale or regularization, and an accept, rollback, re-identify, or stop recommendation. The operator verified the recommendation before any waveform was loaded. A candidate-update round denotes evaluation of a proposed operating waveform. The reported sequence runs from R0 through R30; the response-matrix perturbations and interleaved reference measurements described in Sec.~\ref{sec:feedback} are additional system-identification measurements, not candidate-update rounds. Deterministic scripts performed image fitting, CV calculation, row--column updates, response-matrix assembly, regularized least squares, and CSV serialization.

The model designation follows the contemporaneous project record. Per-turn checkpoint identifiers were not archived, so the manuscript does not assign every round to a frozen backend checkpoint. This limitation prevents a strict model-to-model benchmark and motivates frozen model, prompt, and tool logs in future campaigns.

\section{Response-matrix identification and feedback}
\label{sec:feedback}

The row--column controller used large relaxation factors in the high-error regime and progressively smaller, channel-dependent values as uniformity improved. The rebound at R17 led to rollback to R16 and reduced feedback gains. The CV then approached $2\%$ near R24--R25 in Fig.~\ref{fig:trajectory}(a), motivating the transition to a local response-matrix description.

The local response matrix used the same normalized intensities and logarithmic coordinates as the main text: $M_p=I_p/\bar I$, $u_k=\ln A_k$, and $y_p=\ln M_p$. There are 100 site outputs and 20 tone-amplitude controls, so $J$ has dimensions $100\times20$. Each column of $J$ was measured by a symmetric perturbation,
\begin{equation}
J_{pk}\simeq\frac{\ln M_p(A_k^+)-\ln M_p(A_k^-)}{\ln(1+\epsilon)-\ln(1-\epsilon)}.
\end{equation}
Here $A_k^{\pm}=A_k(1\pm\epsilon)$ specifies the positive and negative perturbations of tone $k$. The first matrix used a perturbation of approximately $\pm0.6\%$. After drift invalidated that local model, a new matrix was measured with $\epsilon=0.004$. Baseline checkpoints cp00, cp10, cp20, cp30, and cp40 were interleaved with the 40 perturbation conditions. Linear interpolation between checkpoints removed slow common drift from the finite differences.

\begin{figure}[!htb]
\centering
\includegraphics[width=0.98\columnwidth]{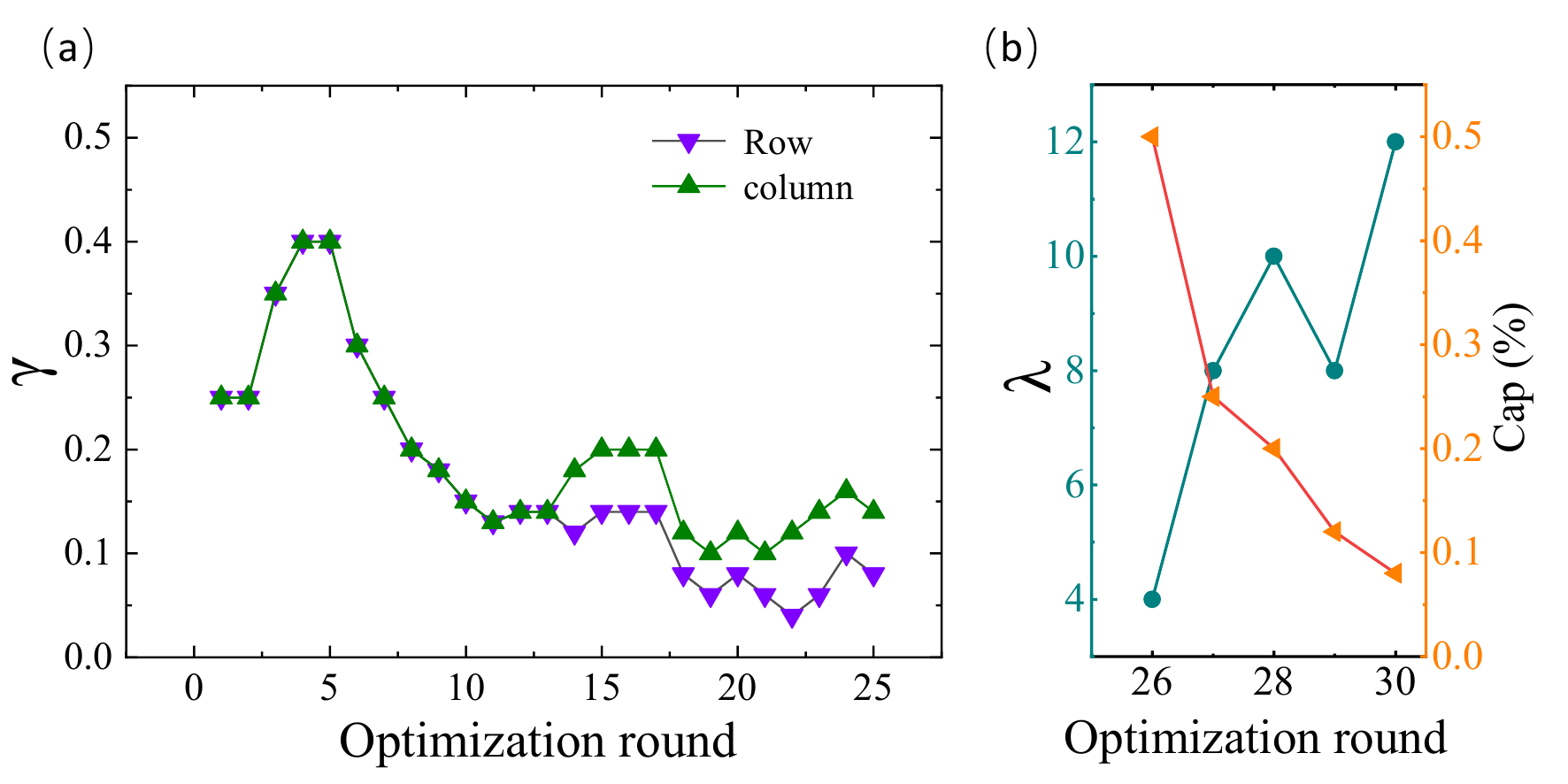}
\caption{Contraction of the experimental update scale. (a) Row and column relaxation factors $\gamma$ during the row--column stage. (b) Regularization parameter $\lambda$ (teal) and per-tone update cap (orange) for successive response-matrix updates. The supplied horizontal-axis numbering is retained. CV values are reported from Fig.~\ref{fig:trajectory}(a).}
\label{fig:schedule}
\end{figure}

The update minimized the Tikhonov objective in Eq.~\ref{eq:ridge}, with the same $\lambda^2$ regularization convention as in the main text, and was then limited by a hard cap on each tone's relative amplitude change. Figure~\ref{fig:schedule}(a) shows the relaxation schedule used by the row and column controller. Figure~\ref{fig:schedule}(b) shows the successive regularization and update-cap settings using its supplied horizontal-axis numbering. These settings illustrate the contraction of the update scale; CV values and state labels in the text follow the trajectory in Fig.~\ref{fig:trajectory}(a).

\end{document}